\documentclass[]{spie}  
\usepackage[dvips]{graphicx}

\title{Gamma-Ray Burst Detection with INTEGRAL/SPI} 

\author{Andreas von Kienlin\supit{a}, Nikolas Arend\supit{a}, Giselher G. Lichti\supit{a},\\ Andrew Strong\supit{a} and Paul Connell\supit{b}
\skiplinehalf
\supit{a}Max-Planck-Institut f\"ur extraterrestrische Physik, Giessenbachstrasse, Garching, Germany \\
\supit{b}The University of Birmingham, Edgbaston, Birmingham, United Kingdom}

\authorinfo{Further author information: (Send correspondence to A. von Kienlin)\\A.v.K.: E-mail: azk@mpe.mpg.de, 
WWW: http://www.mpe.mpg.de/$\sim$azk/azk.html,
Telephone: +49 89 30000 3514\\  
Address: Max-Planck-Institut f\"ur extraterrestrische Physik, Giessenbachstrasse, D-85748 Garching, Germany}

\begin{document} 
  \maketitle 

\begin{abstract}
The spectrometer SPI, one of the two main instruments of the INTEGRAL spacecraft, has strong capabilities in the 
field of Gamma-Ray Burst (GRB) detections. In its 16$^o$ field of view (FoV) SPI is able to trigger and to localize 
GRBs. With its large anticoincidence shield (ACS) of 512 kg of BGO crystals  SPI is able to detect GRBs quasi 
omnidirectionally with a very high sensitivity . The ACS GRB alerts will  provide GRB arrival times with high accuracy 
but with no or very rough positional information. The expected GRB detection rate in SPI's FoV will be one per month 
and for the ACS around 300 per year.  At MPE two SPI software contributions to the real-time INTEGRAL burst-alert 
system (IBAS) at the INTEGRAL science data centre ISDC have been developed. The SPI-ACS branch of IBAS will produce 
burst alerts and light-curves with 50 ms resolution. It is planned to use ACS burst alerts in the 3$^{rd}$ interplanetary 
network. The SPI-FoV branch of IBAS is currently under development at MPE. The system is using the energy and timing 
information of single and multiple events detected by the Germanium-camera of SPI. Using the imaging algorithm 
developed at the University of Birmingham the system is expected to locate strong bursts with an accuracy of better 
than 1$^o$.
\end{abstract}


\section{INTRODUCTION}
\label{sect:intro}  
The launch of INTEGRAL, ESA's next mission devoted to $\gamma$-ray research, is expected soon \cite{Winkler96,Parmar02}. The scheduled launch 
date is October 17, 2002. A Russian PROTON rocket will carry the spacecraft from  the Cosmodrome in 
Baikonur, Kazakhstan on to its higly eccentric orbit with a revolution period around the Earth of 3 days. It has 
a perigee height of 10'000 km and an apogee height of 152'600 km with an inclination of 51.6$^0$. This orbit allows 
long unbroken observations outside the radiation belts. 

After a period of about 3 months, which is dedicated to the commissioning phase and the following inorbit calibration 
and the scientific performance validation , INTEGRAL will start its scientific observation program \cite{Parmar02,Schoenfelder01}. 
The detection and investigation of cosmic gamma-ray bursts (GRBs) is one of the important scientific topics of the INTEGRAL mission.

GRBs are still one of the most fascinating research topics in astrophysics. Until now, since their 
discovery 35 years ago by the Vela satellites in 1967 \cite{Klebesadel73}, this phenomenon is still not totally understood and explained. 
The first major breakthrough in this field was obtained with the BATSE detectors \cite{Fishman89} on NASA's Gamma Ray Observatory GRO mission. 
In the ten years of the GRO mission around 3000 bursts were registered, which showed an isotropic distribution over the entire 
sky, but with a deficiency of weak bursts. It was the Italian/Dutch satellite  BeppoSAX \cite{Boella97} which revealed the cosmological 
nature of GRBs with the identification of the first X-ray afterglow in 1997 \cite{Costa97}, which triggered the first 
successful follow-up observation at optical wavelengths \cite{Paradijs97}.  This finally ruled out the Galactic population models. 
In the meantime an optical afterglow was measured for 25 GRBs, all of them showing redshifts between z=0.36 
and z=4.5 \cite{jcgWWW}. Many new missions, especially built for burst searches or equipped with instruments for the investigation 
of GRBs are planned \cite{Kienlin00}, will be launched soon \cite{Gehrels00} or are already in orbit \cite{Ricker02}.

>From all instruments of the INTEGRAL payload, the two main instruments, the spectrometer SPI \cite{Vedrenne02,Mandrou97,Lichti96}
and the imager IBIS \cite{Ubertini99}, and the 
two monitors, JEM-X \cite{JEM-X} and OMC \cite{OMC}, contributions to the science of GRBs are expected. 
Especially the sensitivity of INTEGRAL's imager and spectrometer above 20 keV will distinguish them from other missions.
Although INTEGRAL is not equipped with an on-board 
burst alert system it has strong capabilites to detect bursts thanks to the on-ground-based INTEGRAL burst-alert system 
IBAS, which was proposed in 1997 \cite{Pedersen97} after the success of the BeppoSAX satellite in rapidly localizing GRBs.   

IBAS is an automatic software system for the near-real time distribution of GRB alerts detected by INTEGRAL to the scientific community. 
The telemetry 
stream of the satellite, which is linked via the mission operation center (MOC) at Darmstadt, Germany, to the INTEGRAL 
science data center (ISDC) \cite{Courvoisier99} at Versoix, Switzerland, will be monitored continuously for the occurrence of GRBs. The  expected 
transmission time of the telemetry from the satellite to ISDC  will be approximately 30 s. Additionally the processing time 
at ISDC  and the time needed to distribute the data to the clients has to be added. In total an observer could expect the 
first alert message within less than 1 min after the onset of the burst. Thus, in the case of a long-duration burst, follow-up 
observations are possible even during the active phase of the burst.
 
The principal structure of the IBAS system is displayed in figure \ref{fig:ibas-structure}, showing the 
programs dedicated to the burst search for the two main instruments IBIS and SPI (two each),  for the anticoincidence system 
ACS of SPI and for JEMX. This report will focus on the three software packages written for the SPI and SPI/ACS. An overview of 
IBAS with details on IBIS are shown in several proceeding reports \cite{Goetz02,Mereghetti01a,Mereghetti01b,Mereghetti00}.

   \begin{figure}[t]
   \begin{center}
   \begin{tabular}{c}
   \includegraphics[width=.7\textwidth]{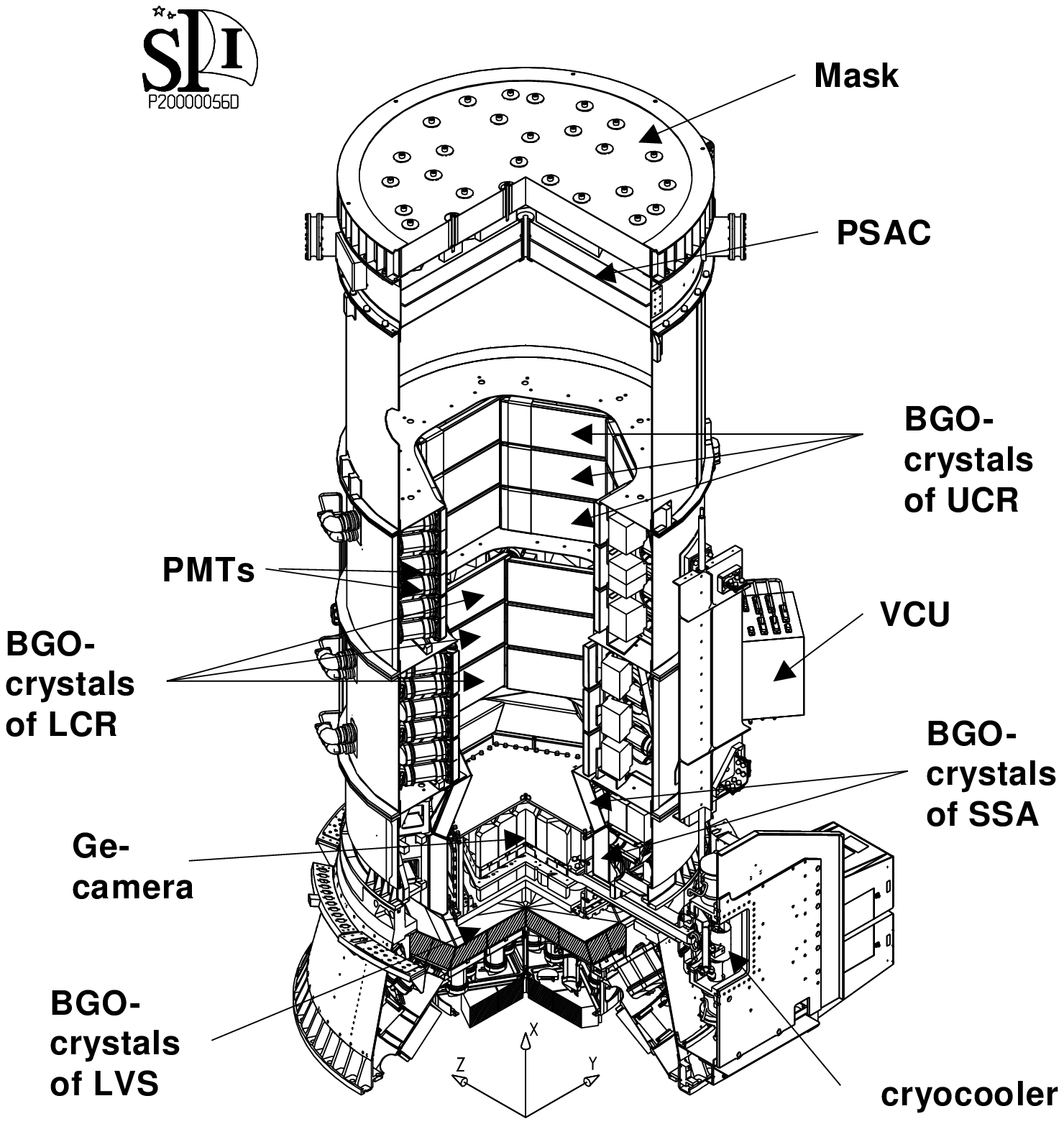}
   \end{tabular}
   \end{center}
   \caption[exampleINTEGRAL Spectrometer SPI] 
   { \label{fig:spi} INTEGRAL Spectrometer SPI.}
   \end{figure} 

\section{The Spectrometer SPI}
The aim of the spectrometer SPI is to perform high-resolution spectroscopy in the energy range between 20 keV and 8 MeV 
\cite{Vedrenne02,Mandrou97,Lichti96}. 
The imaging capability will be good, but it is exceeded by that of the imager IBIS which complements SPI with its 
high imaging resolution, but with less spectroscopic resolving power. So the expected yield and discoveries in the field
of GRB research will be of a different kind for each of the four  INTEGRAL instruments. With SPI new discoveries 
in the spectroscopic regime are possible, supposing that previously-unkown spectral features exist. IBIS will 
provide more precise GRB locations to the community, which is important for the observation of GRB afterglows. The minor 
sensitivity of SPI for GRB detection is compensated by  its larger FoV. So the expected rate/year of detected GRBs is 
about the same for IBIS and SPI. All these quantities are summarized in table \ref{tab:INTperpar}. Also the two monitor instruments 
will help to augment the scientific output for GRBs of the INTEGRAL mission. 

\begin{table}[h]
\caption{This table shows the important performance parameters for the burst detection by the two main INTEGRAL
instruments. PSLA stands for "point-source location accuarcy".} 
\label{tab:INTperpar}
\begin{center}       
\begin{tabular}{|l||l|l|l|} 
\hline
\rule[-1ex]{0pt}{3.5ex}   & IBIS/ISGRI & SPI \\
\hline
\hline
\rule[-1ex]{0pt}{3.5ex}  Energy range & 15 keV - 10 MeV &  20 keV - 8 MeV \\
\hline
\rule[-1ex]{0pt}{3.5ex}  Energy resolution & 9 keV at 100 KeV & 2.3 keV at 1.3 MeV  \\
\hline
\rule[-1ex]{0pt}{3.5ex}  fully coded FoV &  $9^o \times 9^o$ & 16$^o$ corner to corner\\
\hline
\rule[-1ex]{0pt}{3.5ex}  angular resolution & 12' & $2.5^o$\\
\hline
\rule[-1ex]{0pt}{3.5ex}  PSLA & 30'' & $ 10' $ \\
\hline
\rule[-1ex]{0pt}{3.5ex}  Area (cm$^2$) & 2600 & 500 \\
\hline
\rule[-1ex]{0pt}{3.5ex}  bursts / year & $\sim$12 & $\sim$12\\
\hline 
\end{tabular}
\end{center}
\end{table}

Fig. \ref{fig:spi} shows a drawing of the spectrometer SPI.
The camera of SPI, which consists of 19 cooled high-purity germanium detectors, residing in a cryostat,
is shielded on the side walls and rear side by a large anticoincidence shield (ACS).
The field of view (FoV) of the camera is defined by the upper opening of the ACS.  A detailed description of the ACS can
be found in section \ref{chapter-acs}. The imaging
capability of the instrument is attained by a passive-coded mask mounted 1.7 m above the camera. Below
the mask a plastic scintillator anticoincidence (PSAC) allows a reduction
of the 511 keV background, which is mainly generated by particle interactions in the mask.
By observing  a series of nearby pointing directions around the source("dithering") the imaging capability of SPI is 
improved by reducing ambiguity effects.  
For the normal mode of operation two dithering patterns are planned, 
a quadratic $5 \times 5$ and a hexagonal with 7 pointings. The pointing on one grid point will last for about 20-30 min 
with a short 5 min slew between succeeding pointings. Due to the short duration of the bursts, dithering will not improve 
the imaging of SPI for bursts. But during its short duration a burst will be in most cases the brightest $\gamma$-ray source for SPI 
in the whole FoV,  so imaging with SPI is still possible. 
For that it is required that the true source position corresponds to the point in the image with the most-significant detection.
The possibility of observing a burst occurring during a slew, has still to be investigated.
Despite the modest angular resolution of SPI, which is in the order of  2$^o$ , it is possible to locate the direction of 
bursts down to a few arcminutes (see chapter \ref{chapter:senestSPI}). 

\section{Burst detection in SPI's field of view}
\label{GRBsSPIFoV}

The triggering algorithm which will be used for the burst search in SPI's FoV will be
similiar to that used inside IBAS for the IBIS-ISGRI instrument \cite{Goetz02}.
The first algorithm is based only on imaging. It looks for the appearance of new sources by comparing the actual 
image of the sky with those obtained  previously. This algorithm is more sensitive to slowly rising bursts, since the imaging
routine needs some time. The second is expected to be faster, because it monitors the overall event rate in all detectors 
together as function of time. In the case of a trigger the second algorithm uses the imaging algorithm for 
verification. It will test if the count-rate increase is connected with the appearance of a point-like source in the FoV, thus avoiding false triggers
from background variations and other instrumental effects. Both algorithms can run in parallel with multiple instances, using different sets of parameters 
like energy interval or integration time.

Every interaction of a $\gamma$-ray in one of the Ge-detectors is down linked event-by-event, wrapped into packets within the telemetry stream.  
The event-by-event packets contain information on the detector ID which was hit, the energy channel of the recorded pulse-height signal and the time of
interaction, with an accuracy of about 100 $\mu$s. Two kinds of interactions of $\gamma$-rays in the Ge-camera have to be distinguished, in the first case 
a $\gamma$-ray  only interacts within one of the Ge-detectors (single detector events), in the second case the $\gamma$-ray is interacting with 
more than one detector (multiple detector events). The condition for the occurrence of a multiple-detector event is
defined by a coincidence window inside the digital front-end electronics (DFEE) of SPI.  
Both types of multiplicity will be downlinked  with the science telemetry packets, and are used both for the GRB detection.

Additional information on the detector interaction is available through the science house keeping data (HK-data). The rate for each detector is recorded 
on a 1 s basis inside the DFEE and sent to ground, but without energy information. The advantage of these data is that they are always 
available, even in the case of telemetry limitations. For the burst case a 3 Mbit buffer is used on board, and the read out of the buffer could last several minutes,
depending on SPI's actual polling-sequence table (PST). Another application of the science HK Ge-rates could be the case that SPI is working in emergency mode. 
In this mode only the spectra and the multiple-detector events are included in the telemetry.  

\section{The Anticoincidence Shield of SPI}
\label{chapter-acs}

One important part of SPI is its large anticoincidence shield ACS which provides a large effective area 
for the detection of bursts \cite{Kienlin01}, but unfortunately with no or small positional information. But adding ACS to 
the 3$^{\rm rd}$ IPN will avoid this weakness.

The parts of the ACS are also shown in Fig. \ref{fig:spi}.
The ACS  consists of 91 BGO crystals which are arranged in 4 subunits (Fig. \ref{fig:spi}). The units of the upper
veto shield (UVS) are composed of the upper-collimator ring (UCR), the lower-collimator ring (LCR) and
the side-shield assembly (SSA), each containing 18 crystals which are arranged hexagonally around the
cylindrical axis of SPI. The lower veto shield (LVS), consisting of 36 crystals, is assembled
as a hexagonal shell. The thickness of the crystals increases from 16 mm at the top (UCR)
to 50 mm at the bottom (LVS). The total mass of BGO used for the ACS is 512 kg resulting in the
obvious use of the ACS as a burst monitor.

Each BGO crystal of the ACS (with one exception) is viewed by two photomultipliers (PMTs). 
Due to the redundancy concept used for the ACS, each of the 91 front-end electronic boxes
(FEEs) sums the anode signals of two PMTs, which view different BGO crystals
(in most cases neighbouring crystals). This cross strapping of FEEs and BGO
units avoids the loss of a complete BGO crystal if a  single PMT or FEE fails.
It emerges that a disadvantage of this method is an uncertainty in the energy-threshold value of
individual FEEs, caused by different light yields of neighbouring BGO-crystals and different PMT
properties like quantum efficiency and amplification. A result of this is that the threshold
extends over a wide energy range and is not at all sharp.
The actual energy-threshold settings of the ACS will be the result of  a tradeoff between
background reduction and deadtime for the SPI camera.

\section{The ACS as GRB Monitor}
The main task of the ACS as a detector is the generation of a veto signal for the camera 
to suppress charged particles and $\gamma$-rays coming from outside the FoV. 
But the signals from the ACS can also be used for scientific purposes. The ACS housekeeping (HK) data include
the values of the overall veto counter of the veto control unit (VCU) and
the individual ratemeter values of each FEE. Both kinds of HK data are suitable for burst detection.
The count rate of the overall veto counter (ORed veto signals of all 91 FEEs) is sampled every
50 ms. A packet, containing 160 consecutive count rates, will be transmitted every 8 s
to ground. If no gap in the telemetry stream occurs one could have a continuous ACS veto-rate
light curve with 50 ms binning. The measurement time of the individual FEE ratemeter can be
adjusted between 0.1 and 2 s. All 91 FEEs are read out successively in groups of 8 FEEs every
8 s. The readout of all 91 ratemeter values thus needs 96 s. In contrast to the VCU overall veto counter
the individual ratemeter values do not yield a continuous stream. Additionally the
values of different FEE groups are shifted by a time interval of 8 s.
So it is very difficult to derive the burst-arrival direction from these individual counting rates.

\section{ACS as part of the IPN}

The $3^{\rm rd}$ interplanetary networks (IPN) of $\gamma$-ray detectors is in operation since the 
launch of the Ulysses spacecraft in late 1990 \cite{ipn}. Many spacecrafts and experiments have been included
since that time. During the first year of the INTEGRAL mission the IPN will consist of ULYSSES, Mars Odyssey 2001,
Konus-WIND, HETE-II, RHESSI and of course INTEGRAL (SPI/ACS) \cite{Hurley97}. The network will have an excellent 
configuration during this time, due to the large spacecraft separations between Earth, Ulysses and Mars.

An IPN localization of a GRB is obtained by comparing the burst's arrivel times of two spacecrafts ($t_1 - t_2 = \delta T$),
which are separated by a long distance $D$. If the exact positions of the spacecrafts are known, one can localize the burst 
arrival direction to an annulus which spans from  the spaceraft interconnection line with the half-angle $\Theta$ given by the
following expression:
$\cos \Theta = c \cdot \delta T /D$ ($c$: speed of light).
The important quantity, the annulus width is given by 
$\Delta \Theta = c \cdot \sigma ( \delta  T ) /D \sin \Theta $, which is mainly influenced by the uncertainty of the 
arrival-time-difference $ \sigma ( \delta  T ) $ and the spacecraft separation $D$. In the above-mentioned configuration 
of the IPN, with INTEGRAL as near-Earth counterpart, annulli widths in the order of about 1' could be obtained.	
This is due to the high time resolution of the ACS of 50 ms and the spacecraft separations of about 1500 lightseconds.

With a third spacecraft a second annullus can be determined, which has two intersections with the first one. 
This ambiguity can only be solved if one of the spacecrafts provides additionally a rough
positional information. With four or more spacecrafts the burst location is known unambiguously.

\section{The SPI and ACS IBAS software modules}

   \begin{figure}[t]
   \begin{center}
   \begin{tabular}{c}
   \includegraphics[width=0.95\textwidth, angle=0]{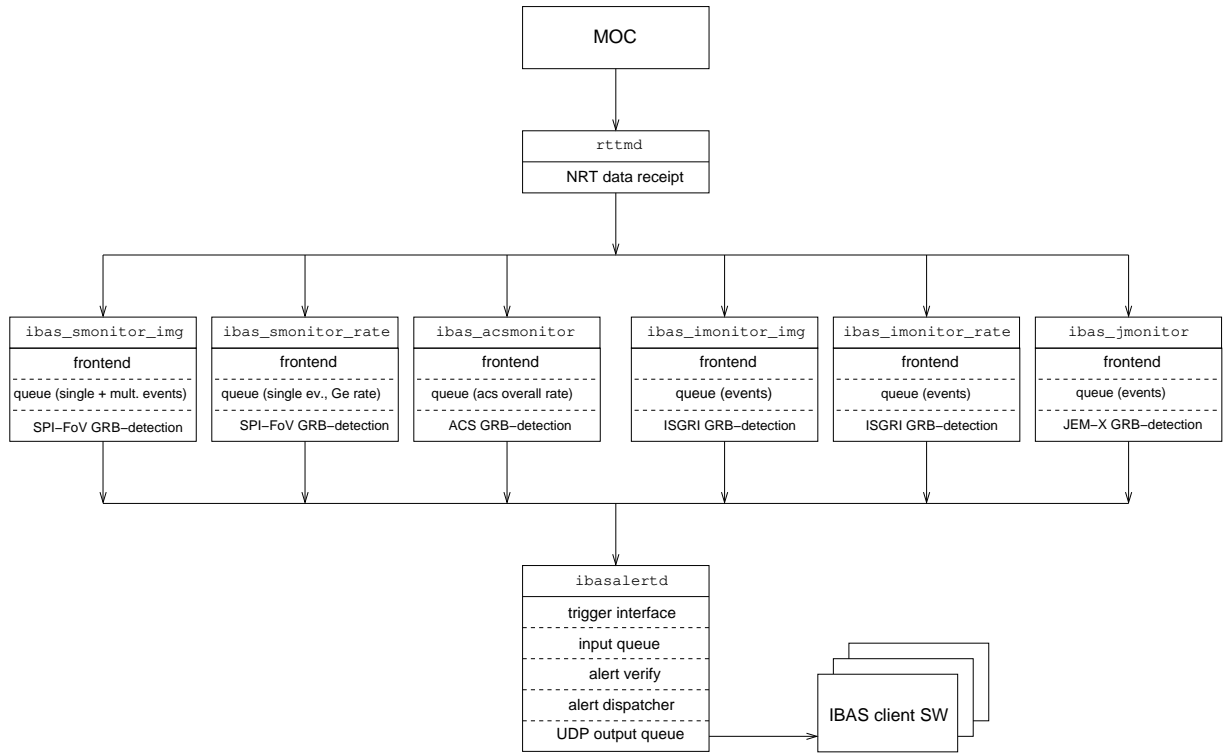}
   \end{tabular}
   \end{center}
   \caption[exampleINTEGRAL Spectrometer SPI] 
   { \label{fig:ibas-structure} Structure of the IBAS. A more detailed description can be found in the "INTEGRAL 
Burst Alert System Architectural Design Document" of the ISDC (IBAS-ADD).}
   \end{figure} 

Figure \ref{fig:ibas-structure} gives a simplified overview on the structure of the whole IBAS system. After the near-real-time data receipt
of the telemetry from MOC, there are several GRB detection programs running in parallel and searching in the telemetry stream for 
the occurrence of a burst. Three of them are responsible for the spectrometer SPI, there are the two programs monitoring the camera events 
({\tt ibas\_smonitor\_img} and {\tt ibas\_smonitor\_rate}) and the monitor program for the ACS overall veto rate ({\tt ibas\_acsmonitor}). 
The other half of the branches monitor the science data of IBIS ISGRI ({\tt ibas\_imonitor\_img} and {\tt ibas\_imonitor\_rate}) 
and JEMX ({\tt ibas\_jmonitor}). The {\tt ibasalertd} program  receives the GRB triggers from the individual branches and 
performs the final alert verification and distribution of the alert messages to subscribed users. 

All IBAS processes are multithreaded applications and run as daemon processes.  
So IBAS is able to perform several subtasks at the same time, with the advantage for the burst search to monitor the telemetry  in 
different energy bands and with different time binning. Each of the GRB detection processes uses the frontend library to read out 
the telemetry packets provided by the data receipt. Then the required instrument data  (Ge single/multiple events, Ge detector count rates, 
ACS veto rates) are decoded from the telemetry packets and put into a time-sorted queue, which is large enough to hold  
several minutes of data. Several trigger-algorithm threads can access the data in the queues, 
in order to detect bursts with different sets of parameter values.

%

The next two subsections present the ACS and SPI-FoV GRB detection programs in more detail.
We begin with the ACS burst-alert system, because this software package has already been delivered to ISDC. The SPI-FoV software is still under development and 
the delivery of the first version is expected soon.

\subsection{The ACS Burst-Alert System}
%

   \begin{figure}[t]
   \begin{center}
   \begin{tabular}{c}
   \includegraphics[width=.65\textwidth, angle=0]{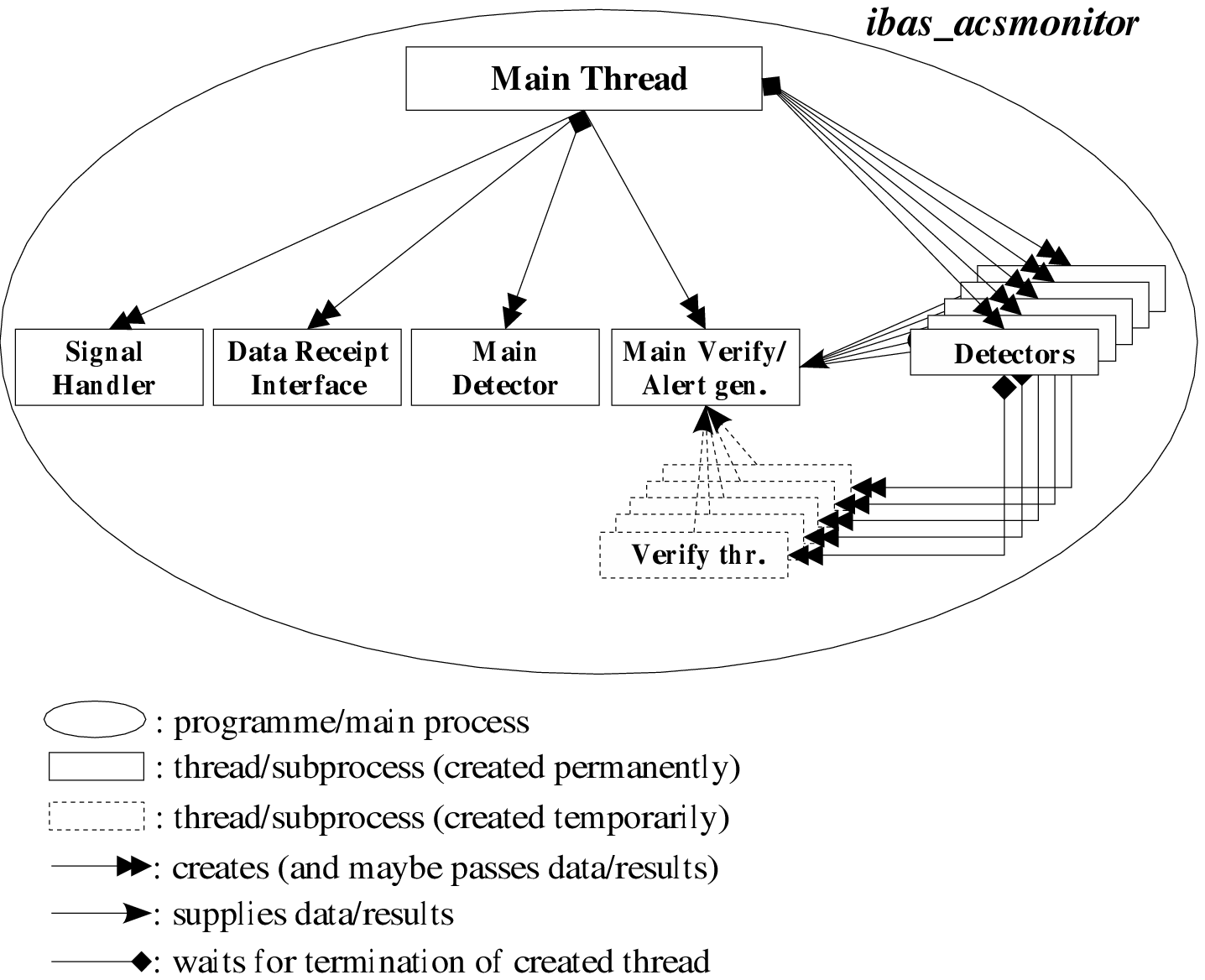}
   \end{tabular}
   \end{center}
   \caption[exampleINTEGRAL Spectrometer SPI] 
   { \label{fig:acsmonitor-thread} Thread structure of the {\tt ibas\_acsmonitor}.}
   \end{figure} 

   \begin{figure}[t]
   \begin{center}
   \begin{tabular}{c}
   \includegraphics[width=.5\textwidth, angle=0]{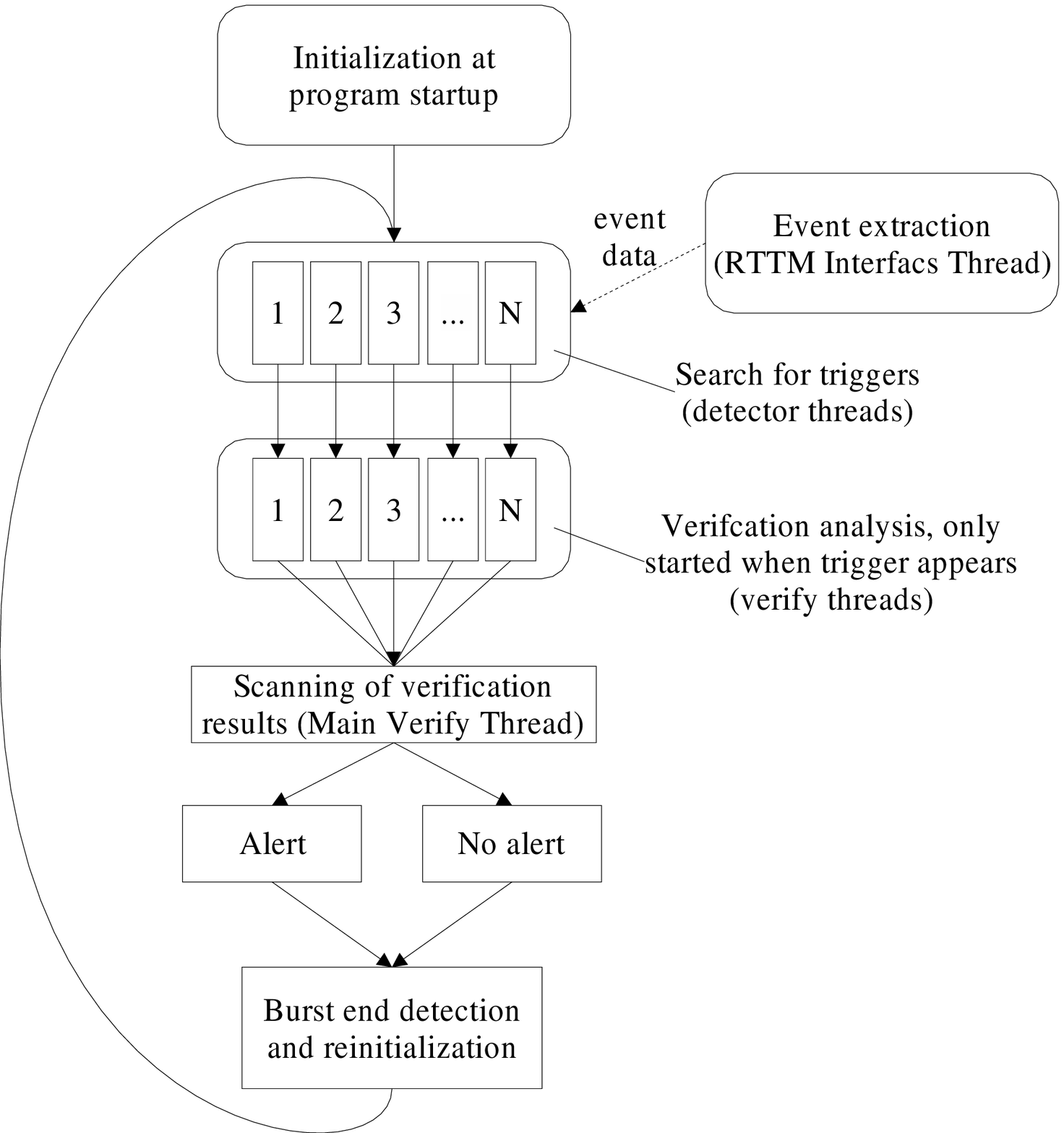}
   \end{tabular}
   \end{center}
   \caption[exampleINTEGRAL Spectrometer SPI] 
   { \label{fig:acsmonitor-steps} Main steps performed in {\tt ibas\_acsmonitor} program.}
   \end{figure} 

The SPI/ACS Burst Alert System {\tt ibas\_acsmonitor} (an earlier version of the program was named {\tt sacs\_grbtd}, "SACS-BAS" \cite{Kienlin01}) 
is one branch of IBAS (see Fig.~\ref{fig:ibas-structure}). 
The trigger algorithm used for {\tt ibas\_acsmonitor} looks
for a significant excess with respect to a running average, comparable to the trigger algorithm 
used for other spacecrafts (e.g. ULYSSES).
%
%
Up to now {\tt ibas\_acsmonitor}  reads only the overall-veto counter values. But it is planned also to include the
read out of ratemeter values of individual FEEs. This will allow a rough
estimation of the GRB arrival direction. 
An accuracy of about $10^{o}$ - $20^{o}$ 
would be enough to distinguish between the two arrival-cone intersections of the interplanetary network (IPN).
The {\tt ibas\_acsmonitor}  trigger algorithm is being tested with generated telemetry data of simulated
burst data plus background. After launch all parameters for the trigger algorithm
and verify criteria will be optimised.

Fig. \ref{fig:acsmonitor-thread}  shows the thread structure of the {\tt ibas\_acsmonitor} program.
The GRB detection threads ("detector"-threads)  search, with different parameter settings (i.e.~timescales, in order to be 
able to trigger on bursts with different temporal behaviour),
for GRBs in SPI ACS light curves. If a significant increase is detected, an analysis 
(trigger verification) is performed by the temporarily created verify threads, to gain for the supression of fault
triggers. In the case of a positive verification result, an alert 
is sent to the alert management system ({\tt ibasalertd}). The number of GRB detection  threads is likewise configurable via a parameter.
Beside the GRB detection and verify threads, {\tt ibas\_acsmonitor}s main program thread creates several other threads 
to perform different tasks: The signal handler thread processes all signals delivered to the {\tt ibas\_acsmonitor} process.
The data receipt interface thread manages the telemetry data transfer, connects (and reconnects, in case of connection
breakdown) to the near-real-time telemetry daemon {\tt rttmd} process via TCP/IP socket. It reads the telemetry data packets, processes 
them and puts data into the local queue. The main detector thread initializes GRB detection threads, including the read in of 
the program parameters. Some checks are performed to guarantee correct variable ranges. The main verify thread watches the 
results of the verify threads and makes the final decision about sending a burst alert to {\tt ibasalertd}. If a burst is 
recognized and an alert must be sent, this thread sets all alert variables defined in alert format.
Fig.~\ref{fig:acsmonitor-steps} gives a rough overview of {\tt ibas\_acsmonitor}s program sequence.

Up to now several verification criteria can be used by the verify threads which are all ORed at the end. The idea of the 
verification tests/criteria is to use characteristic features of the burst light curve to distinguish them from other disturbances.
But the  practicability of each criterium has to be tested after launch. All the criteria are configurable via the parameter file. 
Due to the lack of knowledge of the the burst  direction, the distinction between false and true alert is difficult (e.g. in the case of solar flares).  

The IBAS alert daemon {\tt ibasalertd} has the task to distribute the GRB alert messages to a group of recipients (external users)
which are subscribed to the IBAS-alert distribution list. IBAS alerts are delivered as alert packets via Internet using UDP transport protocol, 
which means alerts can be delivered to the recipients within 1 s. Different types of alert packets are available, with one of them 
especially adapted for the SPI/ACS alerts, containg information on the trigger time in universal time (UT), the significance of the GRB
detection, the spacecraft position and  its attitude (these last two informations are currently not containd in the SPIACS alert
packets. In the next release of the {\tt ibasalertd} this information should be included!). 

Especially for the IPN the burst time history ($\sim 100$~s) is important for the alignment of the light curves obtained from 
different spacecrafts. For this purpose the time history together with the pre-trigger time-history ($\sim 5$~s) will be transmitted 
to the IPN (up to now it is not decided how: via alert package or by making them available on the WWW after a short time).
It is important for the IPN to know the burst-arrival time with  millisecond accuracy. As already shown in \cite{Lichti00} this is 
possible for the ACS overall-counter values.


\subsection{The SPI FoV Burst-Alert System}

The programs for the detection of GRBs inside SPI's FoV comprise the two modules {\tt ibas\_smonitor\_img} and {\tt ibas\_smonitor\_rate} 
(see Fig. \ref{fig:ibas-structure}). Both are currently developed at our institute, and the main structure of the programs has been
taken over from the ACS program {\tt ibas\_acsmonitor} with a subsequent adaption to the new tasks and different data structure. The source-detection imaging 
algorithm/code was delivered by P.~Connell, University of Birmingham. 
The code of the imaging module is part of the {\tt ibas\_smonitor\_ing (\_rate)} packet and the imaging routine is called by the GRB detection threads.


>From the telemetry packets the information on the $\gamma$-ray interaction time, the detector ID, and energy channel are extracted (see also chapter 
\ref{GRBsSPIFoV}). Up to now single and multiple events are analysed. This is sufficient to  do burst imaging in the instrument system. 
If a burst is detected and its position in the instrument system is determined, the sky coordinates can be computed subsequently from 
the pointing information.  The energy channel is used to perform a basic energy selection, but an accurate channel-to-energy
calibration is not required.  The response will be pre-computed for the standard energy channel selections.  The data will be 
accumulated for a time determined by the parameter table and the burst search performed. This is repeated at least at the 8 s
rate of arrival of new data. However the accumulation time can be longer or shorter than this interval; several accumulation times 
can  run in parallel to optimize detection of long and short bursts.  The parameters should be as similar as possible to those for IBIS.  
They include:  accumulation time, significance level, and energy channel range.

For the delivery of the SPI-FoV alerts several burst packets can be distributed by {\tt ibasalertd}. Directly after the GRB detection
a packet containing a first rough estimate of the burst position is sent. Later packets with a refined position will be distributed.

\section{Sensitivity Estimation}
\subsection{GRBs detected in SPI's FoV}
\label{chapter:senestSPI}
%
Simulations and calculations have been performed \cite{Skinner97} in order to investigate SPI's capabilities 
for the burst detection in its field of view. A GEANT Monte-Carlo software, adapted for SPI and a model of the instrument
including a realistic distribution of the matter in the mask, veto-shield and detector assembly was used. By using a strong 
burst from the BATSE catalogue as input, other bursts were generated by scaling flux and duration. For a range of GRB arrival 
directions the detector count rates were simulated and a random background added. Finally these rates were analyzed with the 
independent software system CAPTIF \cite{Connell95} with the  result that SPI exhibits a relatively good performance
for the detection of GRBs.
For bursts with a fluence of $1.1 \times 10^{-6}$~erg/cm$^2$ and a duration of 1 s (only accounting for single detector 
events in the energy range from 20 to 300 keV) the positioning errors inside the FoV ($< 9^o$) will be between 1.5' and 20' and for bursts 
with a larger offset ($< 15^o$) the position error will still be less  than $1^o$. Problems arising from ambiguities and systematic
errors in the determination of the position are more serious towards the edge of the FoV. In summary GRB source locations 
to ~10' are possible over a wide field of view. 
For weak bursts the detection could be problematic, the imperfect imaging together with random fluctuations can lead to a 
spurious peak appearing to be stronger than the true one. The best-estimate position can then be far from the correct location.
The ambiguity problem can occur for bursts with a strength of less than $10^{-7}$~erg~cm$^{-2}$.

As described in section \ref{GRBsSPIFoV}, one method for detecting bursts in the SPI data is the monitoring of the overall
event rate in all detectors together as function of time.  
The minimum-size burst detectable for a backgound rate of ~0.25 cts cm$^{-2}$~s$^{-1}$ in the 50 to 300 keV regime for a 1 s burst,
is about 0.2 photons\cite{Skinner97}, corresponding to $4~ \times 10^{-8}$~erg~cm$^{-2}$. This is about the same as the weakest bursts 
detected by BATSE. 

The rate at which GRBs are detected by an instrument depends on the FoV as well as on the sensitivity. SPI can detect 
bursts at off-axis angles up to at least 15$^o$, so the useful FoV is quite large, ~ 0.2 sr or 680 square degrees 
and consequently the expected detection rate of about 1 burst per month is comparable for IBIS and SPI.

\subsection{ACS effective area and expected GRB rate}
%

   \begin{figure}[t]
   \begin{center}
   \begin{tabular}{c}
   \includegraphics[width=0.8\textwidth,angle=0]{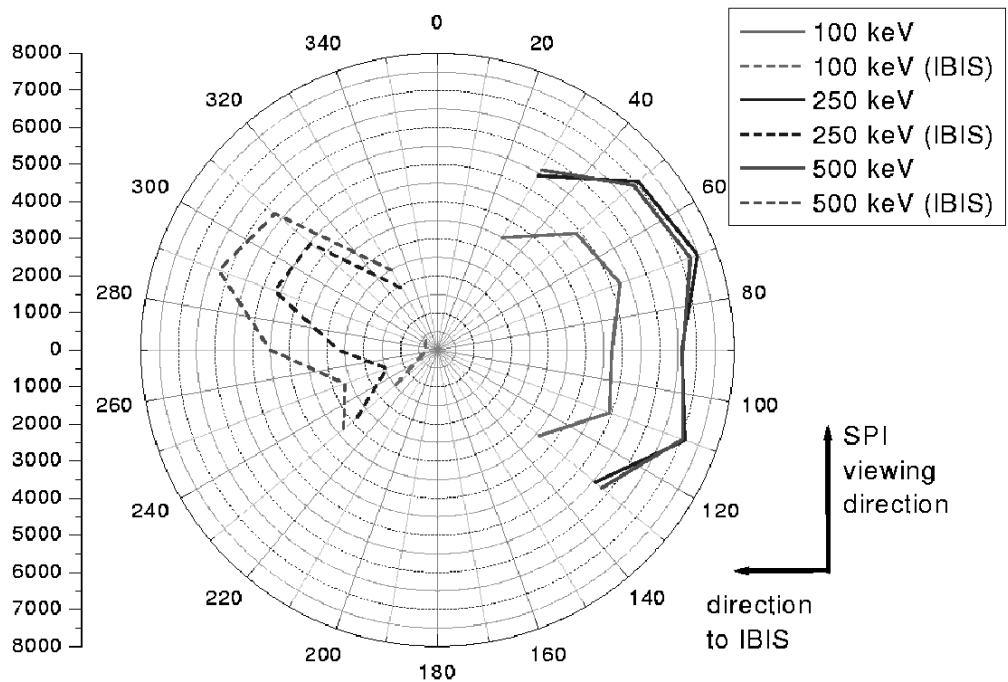}
   \end{tabular}
   \end{center}
   \caption[exampleINTEGRAL Spectrometer SPI] 
   { \label{fig:acs-eff-area} Effective area of the ACS in cm$^2$ as a function of GRB arrival angle and energy. The SPI viewing
direction is 0$^o$. The IBIS instrument occupies the space to the left, leading to a reduced effective area.}
   \end{figure} 

%
An estimation of the expected rate of GRBs detected by ACS has already been given  \cite{Lichti00}.
For an effective area of about 3000~cm$^2$, an ACS background rate between $\sim 80000$~cts/s
and $\sim 160000$~cts/s, an ACS threshold of 80 keV and a time binning of 50 ms the minimum detectable
energy flux is $2 - 2.8 \times 10^{-6} \frac{{\rm erg}}{{\rm cm}^2{\rm sec}}$.
For a time binning of 1 s the sensitivity improves to $5 \times 10^{-7} \frac{{\rm erg}}{{\rm cm}^2{\rm sec}}$.
Using the logN-logP distribution, measured by BATSE and PVO \cite{Fen93} one can derive the number of bursts
which will be observed with the ACS in one year.
The resulting values are $\sim 50$ bursts/year for 50 ms integration time and $\sim 280$ bursts/year for
1 s integration time.
The exact response of the ACS depends on the burst arrival angle, due both to projection effects and to shielding by
neighboring instruments and by the spacecraft structure (see Fig. \ref{fig:acs-eff-area}). These effects together with 
occultation by the Earth on the ACS-"FoV" was not taken into account by the above-mentioned estimation of Lichti 
et al.~2000 \cite{Lichti00}.
The burst intensity can be determined once the infalling direction is known. This is possible only via
Monte-Carlo simulations using the INTEGRAL mass model. Similar simulations have been performed by P. Jean
et al. \cite{Pjean99} in order to determine the sensitivity of the ACS for detection of novae.

\acknowledgments     

We thank Jurek Borkowski from ISDC for his advices and for providing us with specific software routines. 
The SPI/INTEGRAL project is supported by the German "Ministerium f\"ur Bildung und Forschung" through DLR grant 50.OG.9503.0.


\end{document}